\documentclass[aps,prl,twocolumn,superscriptaddress]{revtex4-2}
\usepackage[dvips]{graphicx}
\usepackage{natbib}
\usepackage{datetime}
\usepackage{amssymb}
\usepackage{url}
\usepackage{xcolor}
\usepackage{array}
\newcommand{\olsi}[1]{\,\overline{\!{#1}}} 

\def\be{\begin{eqnarray}}
\def\ee{\end{eqnarray}}

\def\d{{\rm d}}
\def\({\left(}
\def\){\right)}
\def\e{{\rm e}}
\colorlet{red}{black}
\begin{document}
\title{Cooling of gold cluster anions, Au$_N^-$, $N=2-13,15$, 
in a cryogenic ion-beam storage ring}

\author{Klavs Hansen}
\email{KlavsHansen@tju.edu.cn}
\affiliation{
Center for Joint Quantum Studies, Department of Physics, School of Science, 
Tianjin University, 92 Weijin Road, Tianjin 300072, China}

\author{Tian Weihao}
\affiliation{School of Science, Tianjin University, 
92 Weijin Road, Tianjin 300072, China}

\author{Emma K. Anderson}
\affiliation{Department of Physics, Stockholm University, 
AlbaNova, SE-106 91 Stockholm, Sweden}

\author{Mikael Bj{\"o}rkhage}
\affiliation{Department of Physics, Stockholm University, 
AlbaNova, SE-106 91 Stockholm, Sweden}

\author{Henrik Cederquist}
\affiliation{Department of Physics, Stockholm University, 
AlbaNova, SE-106 91 Stockholm, Sweden}

\author{Ji MingChao}
\affiliation{Department of Physics, Stockholm University, 
AlbaNova, SE-106 91 Stockholm, Sweden}

\author{Stefan Ros{\'e}n}
\affiliation{Department of Physics, Stockholm University, 
AlbaNova, SE-106 91 Stockholm, Sweden}

\author{Alice Schmidt-May}
\affiliation{Department of Physics, Stockholm University, 
AlbaNova, SE-106 91 Stockholm, Sweden}

\author{Mark H. Stockett}
\affiliation{Department of Physics, Stockholm University, 
AlbaNova, SE-106 91 Stockholm, Sweden}

\author{Henning Zettergren}
\affiliation{Department of Physics, Stockholm University, 
AlbaNova, SE-106 91 Stockholm, Sweden}

\author{Vitali Zhaunerchyk}
\affiliation{Department of Physics, University of Gothenburg, 
41296 Gothenburg, Sweden}

\author{Henning T. Schmidt}
\affiliation{Department of Physics, Stockholm University, 
AlbaNova, SE-106 91 Stockholm, Sweden}

\date{\today,~\currenttime}
\begin{abstract}
We have measured the spontaneous and photo-induced decays of anionic 
gold clusters, Au$_N^-$, with sizes ranging from $N = 2$ to 13, and 15. 
After production in a sputter ion source, the size-selected clusters 
were stored in the cryogenic electrostatic ion-beam storage ring
DESIREE and their neutralization decays were measured for storage times 
between 0.1 and 100 s. 
The dimer was observed to decay by electron emission in parallel 
to neutral atom emission at long times, analogously to the behavior of 
copper and silver dimers, implying a breakdown of 
the Born-Oppenheimer approximation. 
Radiative cooling is observed for all other cluster sizes.
The decays of clusters $N=3,6,8-13,15$ show only a single 
radiative cooling time.
For $N=6-13$ the cooling times have a strong odd-even oscillation 
with an amplitude that decrease with cluster size, and with the even $N$
having the faster cooling.
We compare our results with previous measurements of radiative cooling 
rates of the corresponding cationic gold clusters, Au$_N^+$, which also show 
an odd-even effect with a similar oscillation amplitude but at orders of magnitude 
shorter time scales, and out of phase with the anions. 
The tetramer and pentamer both show two cooling times,
which we tentatively ascribe to different structural forms at different 
ranges of high angular momenta of the ions in the Au$_4^-$ and Au$_5^-$ beams.  
For Au$_7^-$, the shape of the decay curve suggested that the 
cluster cools by emission of low energy photons. 
The calculated limit on photon energies strongly suggest that cooling is
by vibrational transitions in this case.
For Au$_5^-$, time-resolved studies of photo-induced decays 
were performed to track the evolution of the internal energy distribution.
We conclude that the radiative cooling is dominated 
by sequences of vibrational transitions in the IR.
The laser enhanced neutralization rate of Au$_5^-$ was exponential, 
in contrast to its spontaneous decay rate, 
indicating that the cluster had already been cooled to a very narrow 
internal energy distribution at 120 ms as the 
total (integrated) laser enhanced intensity was independent of the laser 
firing time at later times.
The unimolecularrate constants decreased from 500 s$^{-1}$ when laser excited at 
0.12 s to 40 s$^{-1}$ when laser excited at 0.62 s.
\end{abstract}
\maketitle
\section*{Introduction}\label{intro}

Quantization of electronic energies gives rise to a number of 
characteristic features of metal clusters. 
One of the strongest manifestations of these is odd-even effects, 
which were first observed for the intensity distribution as a function of 
size for small gold clusters by Katakuse et al. 
\cite{Katakuse1985,Katakuse1986}.
The gaps in the electronic excitation spectra associated with this 
effect, and with shell structure, decrease with increasing 
cluster size and ultimately 
give rise to thermally populated electronically excited states 
at some cluster size.
This thermal excitation will cause a strong suppression of both the 
odd-even effect and the shell structure 
(see ref. \cite{HansenCP2020} for an analysis of the thermal suppression 
of the odd-even effect in gold clusters).
It will also cause the appearance of the closely related effect of 
emission of electromagnetic radiation from thermally populated 
electronically excited states, known as recurrent fluorescence (RF) 
\cite{Nitzan1979} or Poincar{\'e} radiation \cite{Leach1987,Leger1988}.

The odd-even intensity effect is particularly strong for gold clusters,
as judged by the abundances in the measured spectra of Katakuse et al.
\cite{Katakuse1985,Katakuse1986}.
Also the recurrent fluorescence is very strong for cationic gold 
clusters \cite{HansenPRA2017-2}.
Intense RF was observed in experiments where cationic gold clusters 
were excited by laser light and unimolecular decays measured in 
a time-of-flight mass spectrometer.
RF implies the presence of low energy optically active electronic 
states and a spectroscopic experiment confirmed the presence of
such absorption features for Au$_{10}^+$ \cite{GreenPRL2021}
(cluster size 10 was the only one measured in that experiment).
The radiative time constants observed in the experiments in 
ref. \cite{HansenPRA2017-2} were very short for clusters 
up to size 9, on the order of microseconds, and showed a very strong 
odd-even oscillation, with radiative cooling times for even-electron 
numbered (i.e. odd $N$) clusters, typically an order of magnitude shorter than 
for the odd electron number clusters, and for both types
orders of magnitude shorter than what can be explained by infrared 
radiation from vibrational transitions.
The correlation of radiative cooling time scales with electron 
number could potentially be ascribed to the higher per-atom 
excitation energy (effective temperature) for the even 
electron number (odd $N$) clusters.
This was found not to be the case, however \cite{HansenJPCC2017}, and 
the main contribution to the strong even-electron radiative emission 
relative to the odd-electron clusters is therefore \textit{not} due to 
their higher stability.
Rather, the systematics is predominantly an intrinsic 
property associated with the number of valence electrons in some other 
way still to be understood. 
A similar conclusion was reached for the opposite question whether the 
observed odd-even abundance oscillations are a radiative cooling effect.
The analysis in \cite{HansenJPCC2017} showed that also this is not the 
case.
The radiative odd-even effect must therefore be considered an 
intrinsic phenomenon for cationic gold clusters.

In addition to the radiative cooling parameters for the 
gold cations, a number of other relevant parameters for the energetics 
of gas phase gold clusters are known. 
These include ionization energies \cite{JackschathBBPC1992}, 
electron affinities \cite{Ho90,Taylor1992,HakkinenJPCA2003},
and cationic dissociation energies \cite{VogelPRL2001,HansenPRA2006}.
Comparing the electron affinities in \cite{Taylor1992} 
and the atomic binding energies in \cite{HansenPRA2006},
it is clear that the two types of binding energies tend to be 
of similar magnitude for a given size, with the reservations 
that the atomic binding energies pertain to cations, and that 
the overlap in cluster size of the two studies is limited.
The lowest activation energy channel determines whether electron 
emission or atomic evaporation dominates in statistical decays, 
modulo the effect of the different frequency factors for loss of 
atoms and of electrons.
The dominant channel sets the energy scale of the 
decaying clusters when decays occur from broad internal energy 
distributions.
We therefore expect the effective temperatures of the present anionic
clusters to 
be similar to those of the cations in the studies summarized in ref. 
\cite{HansenPRA2017-2}.
The question then arises naturally whether the gold cluster 
anions will show RF and if they do, whether they will show the 
same odd-even systematics as the cations. 

Other questions arise as well.
One is whether the effects of the high angular momenta inferred for 
some anionic copper \cite{HansenPRA2017} and silver clusters
\cite{AndersonPRA2018} are also present for gold clusters.
The experimental signature is a decay rate which is a 
sum of two separate curves reflecting different angular momenta populations, 
each described by a power law decay 
initially and modified by a radiatively generated exponential 
suppression at longer times.

A third question is whether the breakdown of the Born-Oppenheimer 
approximation recently seen for the anionic silver and copper 
dimers is also present for the gold dimer anion \cite{AndersonPRL2020,
AndersonPRA2023}.
The breakdown was observed for the copper and silver dimers by 
detection of electron emission in competition with dissociation 
which is the only possible decay channel within the Born-Oppenheimer
description.

\section{Experimental procedure}

The clusters were produced in a cesium sputter ion source with 
currents of $\sim 1$ nA for the smallest species, decreasing with 
size to a fraction of a pA for $N=15$.
The clusters were accelerated to 10 keV ($N=2-9$) or 5 keV 
($N=10-13, 15$), limited by the maximum current in the mass-selection 
magnet.

Prior to injection into the ring, the mass spectrum of the ions 
produced in the source was measured by scanning the magnet. 
The spectrum showed a small but non-negligible 
presence of copper, which originated in the sputter source cathode.
The combination of three atoms of the heavy copper isotope, 
of mass 65 u, is close enough to the mass of a single gold atom,
of mass 197 u, to potentially distort the decay curves by 
contamination.

The amount of such contamination was probed by a combination of 
two procedures.
One was a comparison of the decay spectra recorded after mass 
selection centered at three different positions of nominal gold 
cluster peaks in the magnet scan; one on the low side, one at 
the maximum, and one on the high mass side. 
The two spectra recorded with a setting on the maximum of the 
peak and on its high mass side showed identical decay curves, 
indicative of a pure gold cluster sample. 
There was no sign of hydrides.
In the other procedure, a cathode with a gold surface was used.
A spectrum for the trimer showed a time dependence of the 
decay which was identical to the two high-mass curves of the 
previous test performed with the trimer.
We concluded that these two magnet setting were sufficient to ensure 
beam purity. 
Since the beam current from the gold-surface cathode was not 
as stable as from the standard cathode with a copper holder, the latter 
was used with the gold plug insert for most of the scans.

\begin{figure}
\includegraphics[width=9cm,angle=0]{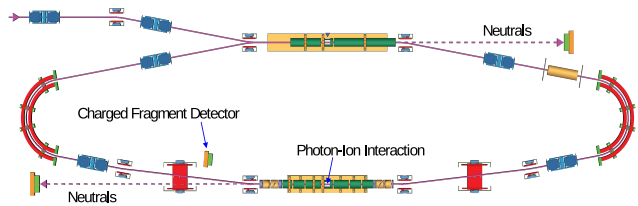}
\caption{A schematic of the ion-beam storage ring used for the present 
experiments on the spontaneous decay of Au$_N^-$ clusters ($N=2-13$ and 15) and 
of laser-induced decay of Au$_5^-$ (see main text).\label{DESIREE}}
\end{figure}

After mass selection, the cluster anions were injected into a 
single ring of the dual ion-beam storage ring DESIREE 
\cite{ThomasRSI2011,SchmidtRSI2013}, shown schematically in 
Fig. \ref{DESIREE}.
The {\color{red} distance} from the source to the midpoint of 
the side where the first fragmentation is detected (the $\ell_{\rm s}$ 
in the appendix) is 6.4 m.
The ions are injected from top left and the neutral products 
from decays of all cluster sizes were detected at the right 
exit side of the top section. 
The experiments with laser excitation applied an OPO 
pulsed laser with pulse durations of around 5 ns and energies of 
typically 1 mJ.
The light was unfocused.
In the Au$_5^-$ laser experiment, the laser beam crossed the ion beam  
at the middle of the bottom section, as
indicated in the figure, and the decay rate was measured with the lower left
neutrals detector.
The dimer decay was measured with two detectors, viz. 
the detector in the upper right corner and the 'fragment detector'.
The top right side detector counts all decays that produce a 
neutral particle and therefore gives the sum of the neutral atom emission 
and the electron detachment signal.
The other ('charged fragment detector') detects only the 
fragments produced in the decay through the atom emission 
channel. 
A comparison of the two detector signals showed the contribution
from the electron emission channel. 
The arrangement is described in more detail in 
refs. \cite{AndersonPRL2020,AndersonPRA2023} for experiments on 
the copper and silver dimer anions.

Storage times varied between 100 ms and 100 s, depending on how
fast the observed decay was quenched by the radiative cooling of 
the ions, if at all.
The revolution times were $83 \sqrt{N}\mu$s for the 10 keV 
beams and correspondingly $\sqrt{2}$ longer for 5 keV 
ions.
For most cluster sizes the ring was filled maximally, which is 
around 90 \%, but for a few of the highest beam currents, which 
were seen at the smaller cluster sizes, the ring was only partly 
filled to avoid detector saturation. 

\begin{figure}
\hspace*{-0.5cm}
\includegraphics[width=10cm,angle=0]{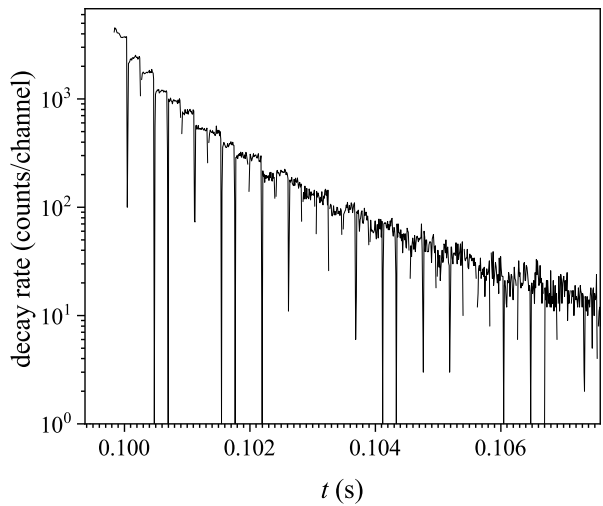}
\caption{The {\color{red} neutralization} count rate of Au$_7^-$ for 
the first thirtysix circulations in the ring, indicated by the plateaus 
of duration $\approx 200 ~\mu$s.
The total storage time was 1 s. 
The {\color{red} time has not been corrected for the} small background 
signal recorded 100 ms before the ions were injected is not subtracted 
from the abscissa in this figure.
The unfilled 10 \% of the ring is seen as the dips in intensity.
\label{zoomRaw}}
\end{figure}

In Figure \ref{zoomRaw} we show part of a raw data spectrum for 
Au$_7^-$ as an example of the measured spontaneous decay.
In the analysis of the spectra, all counts for a turn in the 
ring were summed up. 
Note that all decays in a specific turn can be assigned a single time 
after creation in the source.
Summing over all decays in a single turn does therefore not introduce any 
approximations to the count rate vs. time.
Decay rates are therefore given as counts per turn in the ring.
Counts may be binned over several turns with a bin size that increases 
proportionally with time to place the points equidistantly on 
a logarithmic scale axis. 

In addition to measurements of the spontaneous decay, the 
$N=5$ clusters were exposed to laser light and the 
photon-induced decays measured.
The laser system is EKSPLA NT 342 C, optical parametric oscillator 
tunable wavelength system. 
The light intersected the ions in a crossed beam configuration (see 
Fig. \ref{DESIREE}).
The laser repetition rate was 10 Hz, and the photon energy 2.76 eV.
{\color{red}For comparison the electron affinity is around 3.0 
eV\cite{TaylorJCP1990}.}

{\color{red}The data were recorded after the spontaneous decay of the 
hot clusters produced in the source was quenched by radiative cooling.
The quenching was completed at the first laser firing time of 0.12 s.}
The laser light intensity was kept low enough to avoid depletion of the 
ion beam, and the neutralization ion counts at different laser firing 
times can thus be compared directly.

The measured signal contains two background components besides the 
spontaneous or photo-induced neutralization count.
One is generated by collisions with rest gas molecules.
This is a minor contribution due to the low temperature of the ring 
and the resulting very low pressure \cite{SchmidtRSI2013,BackstromPRL2015}.
The main background contribution is the electronic noise from the 
detectors. 
To determine the detector dark count rate, the first 10\% or 20\% 
of each measurement cycle were recorded before the ions were injected. 
For the two neutrals detectors the dark count rates were around 
5 s$^{-1}$ in the experiment, similar to values seen previously 
in the same detectors.
For the fragment detector it was somewhat higher but not prohibitively 
so. 

\section{Results and analysis}

\subsection{Decays of $N=2$} 

\begin{figure}[ht]
\hspace*{-0.5cm}
\includegraphics[width=10cm,angle=0]{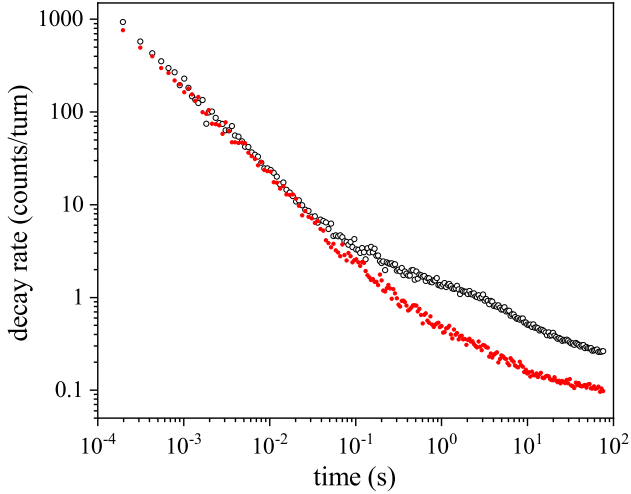}
\caption{The decay rate for the dimer anion vs. time after the 
creation of the clusters in the source; the inclusive detector 
count is given by open black circles and the fragment detector 
count by small filled red circles.
The counts are bunched for statistics, as described 
in the main text.
\label{dimer}}
\end{figure}

In Figure \ref{dimer} we show the decay rate of the Au$_2^-$ dimer 
for the all-inclusive neutral particle detector on the upper right 
side of Fig. \ref{DESIREE} and the fragment detector.
The initial decay channel is completely dominated by the loss of a 
neutral monomer, as seen by the fact that the two curves coincide at 
early times.
The initial decay rate follows a $1/t$ time dependence.
This behavior is expected for decays from an ensemble of clusters
with a broad distribution of rate constants \cite{HansenPRL2001}.
For dimers with a single vibrational degree
of freedom the decays are not thermally activated as in 
the description of usual unimolecular decays.
Instead, the dominant decay at early times (dissociation) is due 
to tunneling out of the angular momentum barrier in the collection 
of the rotationally and vibrationally highly excited clusters created 
in the sputter source \cite{fedor05}.
The power law decay rate arises as a consequence of the 
broad distribution of tunneling rates for {\color{red} this} process.

Around 30 ms, a difference between the two detector curves develops, 
indicating the appearance of a measurable contribution from the 
electron detachment channel, which 
soon after becomes dominant and remains so up to the end of storage 
at 80 s (see Fig. \ref{dimer}).
The onset of a visible contribution from the electron emission 
channel coincides with a departure from the $1/t$ decay rate curve.
The tunneling mechanism is not available for electron emission, 
which instead arises as a result of the breakdown of the 
Born-Oppenheimer approximation \cite{AndersonPRL2020}.
The breakdown is particularly striking for the coinage metal
dimers because the diabatic potential energy curves of the anion and 
the neutral do not cross \cite{Ho90}.

The behavior seen here for the gold dimer is close to identical to 
the ones for the copper and silver dimer anions \cite{AndersonPRL2020,
AndersonPRA2023}, for which decays were measured up to 10 s.
It remains an open question why the three coinage metal dimer 
anions show such similarities for both the onset 
time of a visible contribution from electron emission, and for the 
electron-atom emission branching ratio as a function of time.

\subsection{Decays of $N=3,4,5$}

The decay rates of the clusters Au$_3^-$, Au$_4^-$ and Au$_5^-$ 
are, unlike that of {\color{red} the} larger clusters, not well described
by the combination of a single power 
law and radiative cooling parametrized with a single time constant
{\color{red} (see Eq.(\ref{rate}) below)}.
\begin{figure}
\includegraphics[width=8cm,angle=0]{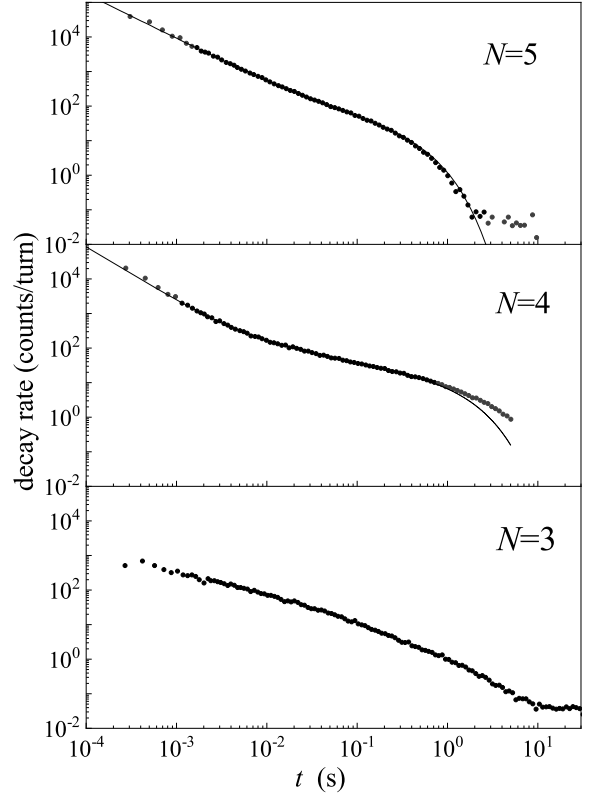}
\caption{The decay rates of clusters (top to bottom) 
$N=5,4,3$.
The lines for the penta- and tetramer are the fitted curves
{\color{red}, using a sum of two radiatively quenched decay rates.}
The trimer curve did not give a good fit.
All spectra are measured with the all-inclusive neutral detector.
\label{fig345}}
\end{figure}
The decay curves for these three clusters, shown in Fig. \ref{fig345}, 
are very similar to those seen for silver and copper anions of the same sizes 
(see refs. \cite{HansenPRA2017,AndersonPRA2018}).
A significant part of the analysis of the small silver and copper 
clusters apply with little change to the three gold cluster anions.
Briefly, the source will produce the gold clusters with a range of states 
with high rotational excitations. 
A distribution of angular momenta can give rise to two, or a few, 
different geometries of the clusters that are realized at different angular 
momenta, as discussed in detail in \cite{HansenPRA2017}.
Conservation of angular momentum protects such conformers from free 
interconversion and causes the ions to decay as an ensemble of different 
species with different properties. 
Different geometric structures will in general have different radiative 
cooling rates, i.e., two or more radiation constants may be seen in the 
measured decay rate.
We suggest that the presence of two such populations in the beam
causes the structure in the decay rates of $N=4,5$, with their first relatively 
steep decrease followed by a slower decrease for the intermediate times between 
10 and 100 ms.
This behavior is most clearly seen for the tetramer ($N=4$) but is
also present in weaker form for the pentamer ($N=5$). 
The suggestion is borne out by the fitted powers of the initial 
decay, which (numerically) exceed unity ({\color{red} -1.30 for $N=5$ and
-1.58 for $N=4$}) and which is then understood as the residual of a 
radiative cooling component which has almost quenched at the beginning of 
the measurement time range of a couple of hundreds microseconds. 
The values of the fitted powers reflect this initial cooling.

The trimer decay rate does not show the intermediate flattening 
of the decay curve.
Another difference from the tetramer and the pentamer is that the 
absolute value of the power on the initial decay is significantly 
below unity {\color{red} see Table \ref{table1}}.
The only possible explanation we can give for this is that 
the trimer decays in a manner similar to the dimer, by electron 
emission and for similar reasons, up to $\sim$ 10 ms, after
which exponential quenching by radiative cooling sets in
(the flat part of the spectrum after 10 s is background counts).
To wit, the dimer decay rate flattens out at long times and 
appears to mimic a power law decay with a small power.
Although the similarities are suggestive, the mechanisms behind 
this flattening is still unclear.

\subsection{Decays of $N=6-13,15$}

\begin{figure}[ht]
\hspace*{-1.5cm}
\includegraphics[width=11cm,angle=0]{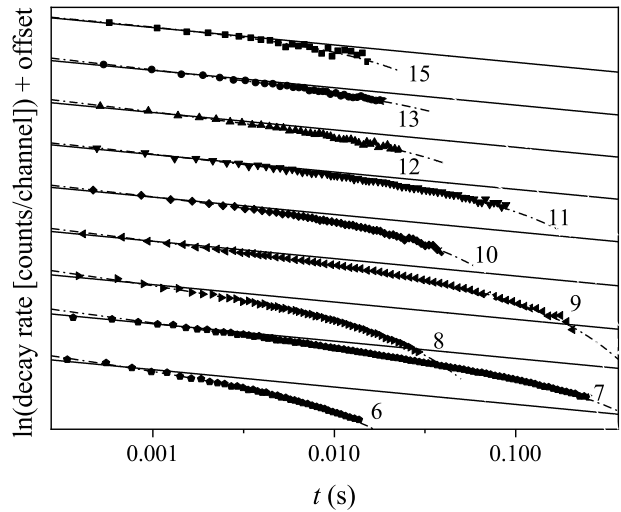}
\caption{The measured decay rates for clusters $N=6-13,15$
vs. time, $t$, after the creation of the clusters in the source.
The curves have been offset by a constant amount for display 
purposes.
The ordinate scale is provided by the straight lines that
are proportional to $1/t$.
\label{6-15fitplot}}
\end{figure}

In Figure \ref{6-15fitplot} we show the measured spontaneous decay 
curves for $N=6-13,15$.
As for the smaller clusters, the production of the clusters in the 
source leaves them highly excited at the time of injection into the ring. 
The curves have been fitted with Eq.(\ref{rate})
(also $N=7$; see below for a refit with Eq.(\ref{smallph}).)

\begin{figure}[ht]
\hspace*{-0.5cm}
\includegraphics[width=9cm,angle=0]{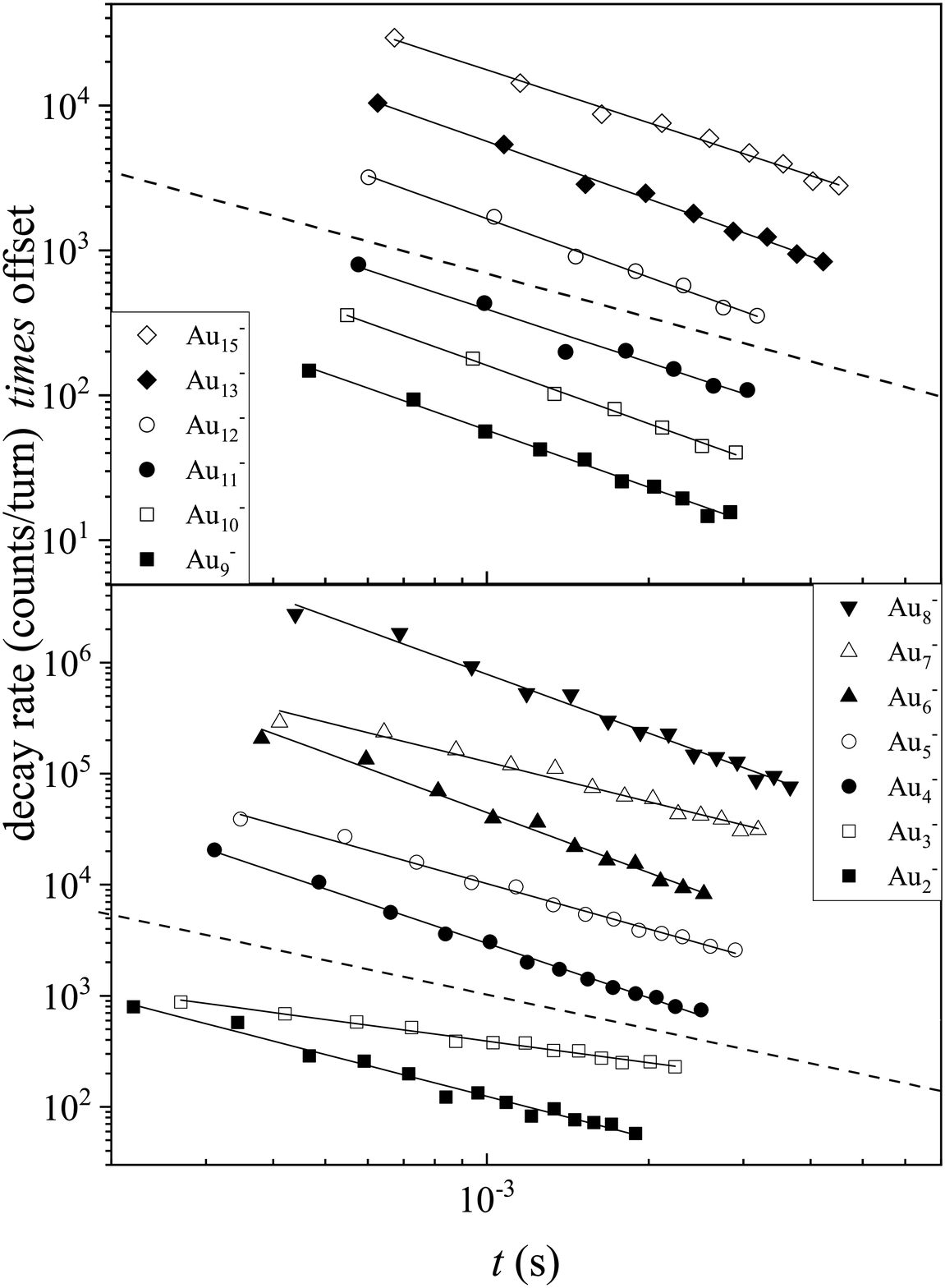}
\caption{Fits of the initial decay rates for all clusters
Au$_N^-$, N=2-13, 15.
The dashed line is the $1/t$ decay multiplied by an 
arbitrary constant.\label{powerlawfits}}
\end{figure}

All sizes shown in Fig. \ref{6-15fitplot} decay initially 
with close to the expected $t^{-1}$ law rates, but 
with some powers in $t^{-p}$ differing significantly from $p=1$.
A zoomed view of the initial decays is given in 
Fig. \ref{powerlawfits}.
The fitted powers are given in Table \ref{table1}

At later times the decays quench and approach a quasi-exponential 
form for the decay rate.
This is ascribed to radiative energy dissipation.
An alternative suggestion is the freezing-in of vibrational degrees 
of the product cluster, as seen for SF$_6^-$ \cite{MenkPRA2014}. 
However, estimates of the level densities rule this out for the 
present study of gold cluster anions and we will disregard this 
mechanism. 

The radiative time scales are clearly dependent on cluster size, but 
also the fitted powers of the initial decay show some 
variation with size. 
In the simplest form the quenching of the neutralization rate is 
described by the exponential factor in the expression
\be
\label{rate}
R \propto t^{-p}\e^{-k_{\rm r}t} \equiv  t^{-p}\e^{-t/\tau} .
\ee
This equation pertains to the emission of large energy photons
with rate constants $k_{\rm r}$; large energy in the sense 
that the emission of a single photon quenches any further unimolecular 
decay.
Eq.(\ref{rate}) is the expression used for the fits for most cluster sizes 
here.
The alternative expression, given below, describes the time constant for 
photon energies that are so small that their effect must be described as 
continuous cooling of the cluster.
The spontaneous decay of Au$_7^-$ is described better by that form.
A thorough discussion of the area of applicability of the large and small 
photon energy quenching is given in Ref. \cite{FerrariIRPC2019}.
We show in Table \ref{table1} the powers $p$ and time constants $\tau$ from 
fits with Eq.(\ref{rate}), including a fit for the heptamer for a relatively 
short time range.
\begin{center}
\begin{table}
\caption{Powers and time constants as fitted with Eq.(\ref{rate}). 
The entries for $N=7$ refer to fits at relatively short time.
Standard deviations of $p$ and $\tau$ are given as $\sigma_p$ 
and $\sigma_{\tau}$, respectively.}
\begin{tabular}{c c c c c}
\hline\label{table1}
 $N$ & $~~~p~~~$ & $~~~\sigma_p~$ & $\tau$ (ms) & 
$\sigma_{\tau}$ (ms)\\
\hline

 3 & 0.79 & 0.01 & 2800 & 200 \\

 6 & 1.49 & 0.04 & 4.5 & 0.2 \\ 

 7 & 1.44 & 0.01 & 120 & 12\\

 8 &1.48 & 0.02 & 7.0 & 0.3 \\
 9 & 1.31 & 0.01 & 50 & 1 \\
 
 10 & 1.15 & 0.02 & 13.0 & 0.7\\
 
 11 & 1.15 & 0.02 & 42 & 3 \\
 
 12 & 1.22 & 0.06 & 17 & 3 \\
 
 13 & 1.18 & 0.05 & 24 &7 \\
 
 15 & 1.16 & 0.07 & 17 & 4\\
\hline
\end{tabular}
\end{table}
\end{center}

\begin{figure}
\hspace*{-0.5cm}
\includegraphics[width=10cm,angle=0]{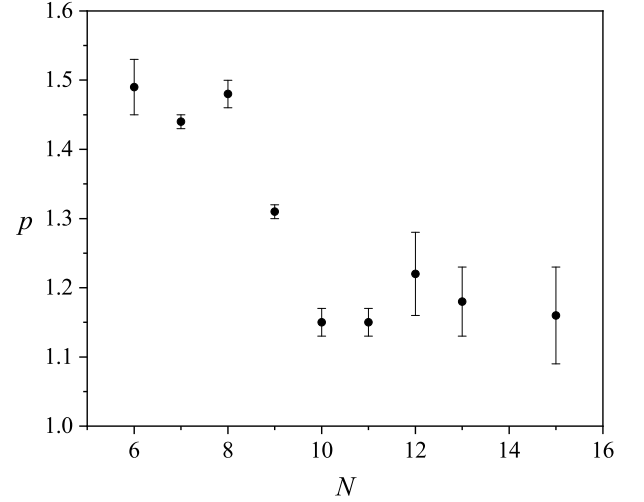}
\caption{The fitted values of the power in the initial 
power law decay $p$, listed in Table \ref{table1}.
The decays were single component curves for the clusters
shown, $N=6-13,15$.
The error bars give the one standard deviation fitted values.
\label{p}}
\end{figure}

In Figure \ref{p} we show the fitted values of $p$ for $N\geq 6$.
There is a clear deviation from a unit power, $p=1$, which decreases 
with size and reaches a value a little above unity for the larger cluster sizes.
The value of $p$ is expected to deviate slightly from unity, mainly due 
to the effect of the finite heat capacity of the clusters.
The expression is \cite{book}
\be
p = 1 + \delta + 2\delta \frac{\e^{-\delta \ln{\omega t}}}
{1-\e^{-\delta \ln (\omega t)}},
\ee
with $\delta \equiv 1/(C_{\rm v}/k_{\rm B}-1)$, $C_{\rm v}$ the canonical 
(vibrational) heat capacity, where $t$ is the observation 
time and $\omega$ the frequency factor of the observed unimolecular decay, 
thermionic emission or atomic evaporation, as it may be.
The expression reduces to $p \approx 1 + 2/\ln(\omega t)$ for 
large heat capacities.
The calculated values for the clusters $N=10-15$ are 1.08 to 1.09 
for $\ln (\omega t) = 25$ and an effective heat capacity of 
$(3N-7)k_{\rm B}$, corresponding to thermionic emission from harmonic 
oscillator clusters ($3N-7$ is the relevant value for 
a microcanonical ensemble of harmonic oscillators, 
see \cite{AndersenJCP2001}).
The observed values for $N\geq 10$ are a little higher than this estimate but 
not unreasonably so, considering that
the heat capacities are not precisely known.
For the smaller clusters, $N=6-9$, the observed values of 1.3 to 1.5 are 
too large to be explained in that way.

The explanation for the high values that we find most plausible
is that the measured power is masking the residual of a radiative cooling 
component which occurs during transit from the source to the ring.
This explanation requires the co-existence of isomer populations in the beam, 
similar to those inferred previously for rotationally highly excited species 
for the copper and silver anion clusters \cite{HansenPRA2017,AndersonPRA2018}
and as discussed above for Au$_4^-$ and Au$_5^-$.
Further experiments are required to clarify the picture.

\begin{figure}
\hspace*{-0.5cm}
\includegraphics[width=10cm,angle=0]{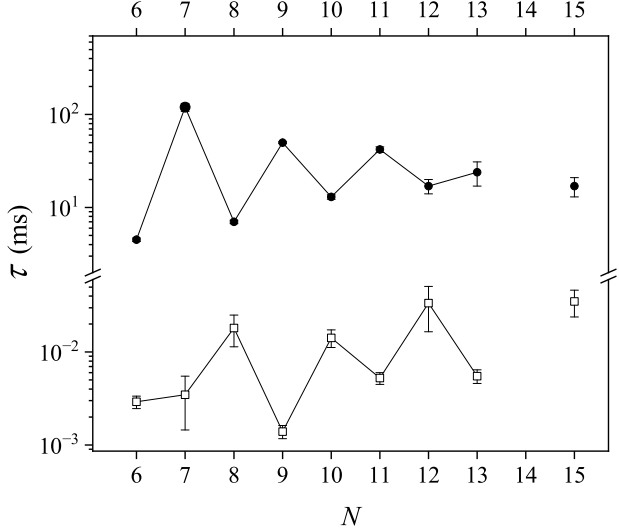}
\caption{The radiative cooling constants measured here for 
the gold anion clusters $N=6-13,15$ (filled circles).
The open squares show the time constants for cationic gold clusters
(The values are slightly shifted from the values in \cite{HansenPRA2017-2}
due to the inclusion of an additional data set).
\label{tau6-15}}
\end{figure}

In Figure \ref{tau6-15} we show the radiative cooling time constant, $\tau$, 
for Au$_N^-$, $N\geq 6$, as filled circles.
The data show a striking odd-even effect, similar in magnitude to that 
of the \textit{cations}, although with two important differences.
One is that the time constants are three to four orders of magnitude 
larger than those for the cations \cite{HansenPRA2017-2}, 
shown as open circles.
The other difference is that the effect have opposite phases for anions 
and cations.

The presence of the odd-even effect in the radiative cooling of the anions
has features that give conflicting indications of its origin.
On one hand, the bare existence of odd-even oscillations as a function of
the number of electrons indicate an electronic origin, similar to the
odd-even effect in abundances and in the (cationic) cluster binding energies.
The electron spin degeneracy of single electrons summarizes this 
oscillationary behavior.
The odd-even staggering  is then the result of electronic structure.
With respect to radiative cooling, this indicates RF emission, at least 
for the even-$N$ clusters.
Likewise, the similar amplitude in the size-to-size oscillations 
for anions and cations suggests an electronic origin of the cooling.
Such variations in oscillator strengths in vibrational emission are not seen 
in other systems to our knowledge.

On the other hand, the average emission rates are rather small for 
RF radiation.
And furthermore, the periodicity has, as noted, the opposite phase
to the cationic case.
This is not expected for electronic transitions because the difference in charge 
state of cations and anions is two, corresponding to a full period of the 
oscillations.

We will tentatively conclude that the radiation involves vibrational 
cooling in some form, although contributions from electronic transitions 
can also not be excluded.
One possibility is that the even-electron species have access to low-lying 
electronic excitations that are absent for odd-electron clusters. 

\subsection{$N=7$}

The fitted value for the power and radiatively cooling 
constant for clusters $N=7$ that are given in Table \ref{table1} 
refer to relatively short times after production in the source. 
Including longer times yielded systematically different 
parameters of the fit function $t^{-p} \exp(-t/\tau)$.
To understand this better we will analyze the decay of this cluster 
in more detail.
One step is to consider the expression for cooling through emission 
of photons with small energies for which the 
neutralization rate becomes \cite{FerrariIRPC2019}
\be
\label{smallph}
R \propto 1/\(\exp(t/\tau)-1\),
\ee
The expression describes the situation where the clusters retain enough 
energy to decay even after emission of one or several photons.
It is derived by an expansion of the logarithm of the rate constant, 
$k$, with respect to energy and by assuming that the emitted power, 
$-\d E/ \d t$, is time independent \cite{HansenJCP1996,FerrariIRPC2019}:
\be\label{lowphotonEs}
1/\tau = \frac{\d \ln k}{\d E} \(-\frac{\d E}{\d t}\).
\ee
A simple derivation of the expression in Eq.\ref{smallph}
is found in ref. \cite{FerrariIRPC2019}.
For completeness we reproduce the calculation here:
With a constant emitted power, the rate constant for a cluster of initial 
energy $E$ varies to first order in time approximately as 
\be
k(E,t) = k(E,0) \e^{-{\rm w}t},
\ee
with 
\be
{\rm w} \equiv {\color{red}-}\frac{\d \ln k}{\d t}.
\ee
{\color{red}The derivative is taken at zero time at the energy, $E$, that 
corresponds to the one for which
the rate constant is equal to the reciprocal radiation time, $k(E)= 1/{\rm w}$,
similar to the procedure used in ref. \cite{andersen1996}.}
The exponential arises because of the strong variation of $k$ with energy,
and does not require any other feature of the energy dependence. 
Calculating the observable total decay rate, $R(t)$, with due consideration 
to the depletion prior to $t$ gives
\be
&&R(t)\propto\\\nonumber
&&\int_0^{\infty} g(E) 
\exp\(-\frac{k(E,0)}{\rm w}\(1-\e^{-{\rm w}t}\) \) k(E,0)\e^{-{\rm w}t} \d E
\ee
where $g(E)$ is the initial {\color{red} distribution of excitation energies, $E$, 
of the clusters.}
Up to multiplicative factors that are time independent to a good approximation, 
the time dependence of this integral is identical to the one that gives the 
non-radiative power law decay after the substitution 
$t \rightarrow (1-\exp\(-{\rm w}t\))/{\rm w}$.
In the applications it should be noted that the expression is derived with
a first order expansion and it will therefore not 
cover decays at all times but it is a decent approximation and it 
differs in an observable manner from the large photon energy limit 
of Eq.(\ref{rate}).

In order not to bias the fit, an expression for the parallel 
emission of both small and large photons is used:
\be
\label{generalR}
R \propto t^{-p'} \frac{\exp(-t/\tau_1)}{\exp(t/\tau_2)-1}
\ee
Although the fit will show that $\tau_1$ is effectively
infinite, i.e., that there is an unobservable amount of large photon 
energy cooling, clearly the full expression (Eq.(\ref{generalR})) is 
required to show this.
We find it useful to include it also for future reference.
The value of $p'$ is the correction to the 
$p=1$ power law for non-ideal behavior, i.e., $p' = p-1$, as can be 
seen if the exponential in the denominator is expanded for times short 
relative to $\tau_2$.
The interpretation of Eq.\ref{generalR} is interesting.
The equation expresses the two reasons the neutralization decays can 
be suppressed by radiative cooling:
One is the factor giving the large photon suppression,
$\exp(-t/\tau_1)$.
This has the effect that the \textit{population} of the decaying particles 
decays exponentially.
Or more correctly, to be converted into clusters with energies that 
are so low that the ions wil remain intact.
The other factor, due to emission of small energy photons, $1/(\exp(t/\tau_2)-1)$,
causes instead the unimolecular \textit{rate constants} to decrease exponentially.
See ref. \cite{FerrariIRPC2019} for details and ref. \cite{HansenPRA2020}
for an example of parallel large and small photon energy radiative 
channels for C$_{60}^-$.

\begin{figure}
\hspace{-0.5cm}
\includegraphics[width=10cm,angle=0]{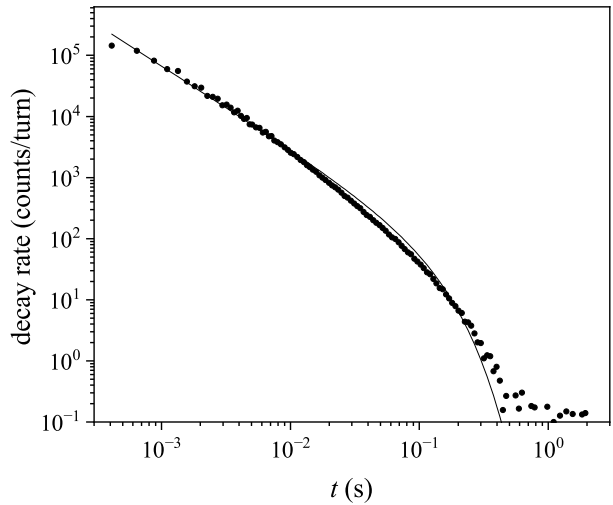}
\caption{Experimental decay rate (points) and the function 
fitted with Eq.(\ref{generalR}) (line) for Au$_7^-$.\label{N7fit}}
\end{figure}

In Figure \ref{N7fit} we show a fit of the decay curve with the
modified, general cooling rate expression in Eq.(\ref{generalR}). 
The curve is best fitted with $p' = 0.35$, $\tau_2 = 60$ ms, 
$\tau_1=\infty$, compared to the values $p'=0.44$, $\tau_1 = 120$ ms
from the fit above to Eq.(\ref{rate}) where instead $\tau_2$ 
was constrained to be infinite.

With some assumptions, the value of $\tau_2$ can be used to find the 
emitted power. 
The rate constant, $k$, describing electron emission of 
a cluster with electron affinity EA and a level density of $(E + E_0)^{s-1}$
is,
\be
\label{harmonic}
k(E) = \omega \(\frac{E+E_0 - {\rm EA}}{E+E_0}\)^{s-1}.
\ee
The energy $E_0$ is an offset in the caloric curve, and $s$ is the 
canonical heat capacity in units of Boltzmann's constant.
The value of $E_0$ does not enter the final result and is only included 
to give some generality to the expression.
Assuming a frequency factor of 10$^{14}$ s$^{-1}$ for thermionic emission
(thermal electron emission), that the cluster is described 
by harmonic vibrations, {\color{red} and given} the electron affinity of 3.5 
eV \cite{Taylor1992}, we can calculate the missing term in 
Eq.(\ref{lowphotonEs}) and express the emitted power by the measured 
time constant as
\be
-\frac{\d E}{\d t} = \tau_2^{-1} \(\frac{\d \ln k}{\d E}\)^{-1}
= 0.57~{\rm eV/s}.
\ee
The energy content differs from threshold energy for the process.
This \textit{kinetic shift} of 12\% has been included in the value given.

If the decay curve is to be represented by the small photon energy limit,
the emitted photons must be smaller than the width of the energy
distribution of the clusters that emit the electrons.
Energy-resolved decay rates, $r(E)$, occur in the general case according to 
\be\label{gaussian0}
r(E)\d E \propto \e^{-k(E)t}k(E) \d E.
\ee
This is a strongly peaked distribution as a function of energy, 
We can calculate an estimate of the width of 
the energy distribution by approximating the right hand side of 
Eq.(\ref{gaussian0}) with a Gaussian, 
\be
r(E) \propto \exp (-(E-\olsi{E})^2/\sigma_E^2 ).
\ee
We estimate the standard deviation to
\be
\label{width}
\sigma_E \approx \frac{\sqrt{2}}{s-1}{\rm EA} \(\omega t\)^{-1/(s-1)}.
\ee
This then gives for the heptamer
\be
\sigma_E = 0.043 ~{\rm eV}.
\ee 
Using that estimate for the highest possible photon energy
for which the fit is applicable, $h\nu = 0.043$ eV, gives a photon 
emission rate constant 
of no less than 0.57/0.043 s$^{-1}$ = 13 s$^{-1}$.
Although this lower limit for the emission constant is easy to fulfill
for both types of cooling, vibrational and RF, it is 
consistent with vibrational cooling, and with the upper limit of the 
photon energy of 0.043 eV, it strongly suggests that 
the cooling is vibrational in the case of Au$_7^-$.

\subsection{$N=5$ photo-excitation}

The photo-induced decay of Au$_5^-$ was measured with 450 nm 
(2.76 eV) light from 0.12 s to 0.92 s, in steps of 100 ms.
{\color{red} The decays were measured on time scales from hundred microsecond
to several milliseconds after laser excitation.}
The energies sampled are therefore 2.76 eV below {\color{red} the energy 
corresponding to decay time constants on the several} hundred 
microsecond time scale.

\begin{figure}
\includegraphics[width=10cm,angle=0]{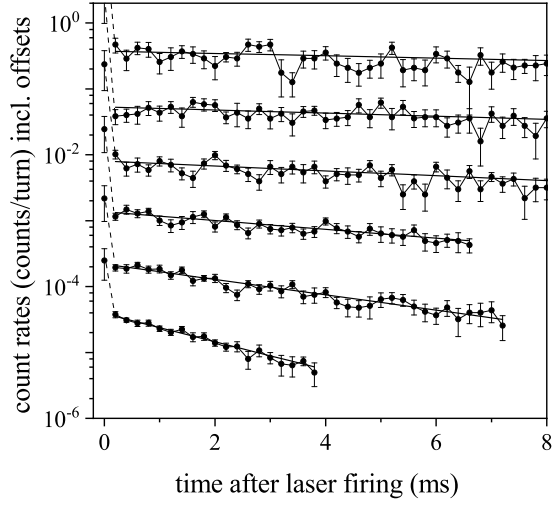}
\caption{The measured photo-enhanced signal of Au$_5^-$, 
turn by turn after excitation. 
The curves (displaced on the ordinate with a constant factor
between each) are recorded at laser firing times of 
(bottom to top), 0.12 s, 0.22 s, etc up to 0.62 s. 
Error bars are statistical.
The curves for the laser firing times 0.72, 0.82, and 0.92 s 
were similar but too flat to yield time constants.\label{Au5laser}}
\end{figure}

\begin{figure}
\hspace*{-0.7cm}
\includegraphics[width=10cm,angle=0]{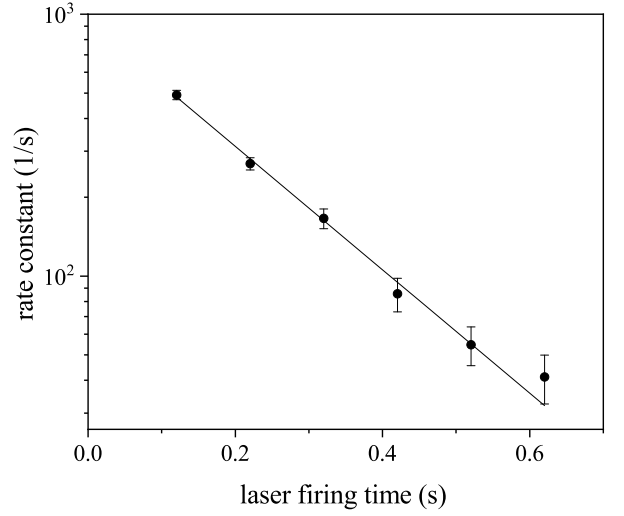}
\caption{The fitted rate constants for the time decay of the photon 
enhanced textcolor{red}{electron emission} signals for Au$_5^-$ shown 
in Fig. \ref{Au5laser}.
Error bars are calculated by error propagation.
The fitted straight line shown has a slope of 
4.96 s$^{-1}$.
\label{Au5const}}
\end{figure}

In Fig. \ref{Au5laser} we show the decays as functions of time 
for each turn in the ring after the laser pulse for different laser 
firing times between 0.12 s and 0.62 s.
Excluding the first peaks, the decays are very well represented 
by single exponentials.
The first peak is dominated by fast decays caused by absorption of 
two or more photons.
The decays during the following turns in the ring must be single photon
absorption to be consistent with a single exponential decay.
(A mixture of single and double photon decays will give curves composed 
of at least two exponentials.)
The fitted photon emission time constants are shown in 
Fig. \ref{Au5const} vs the laser firing time. 
There is a clear and smooth decrease of the time constants with 
laser firing time, corresponding to a rate constant that varies
as $k(t) =925 ~{\rm s}^{-1}\exp(-4.96 {\rm s}^{-1}t)$.
The total integrated intensities of the six spectra are identical
within the uncertainties.
The integrated intensity of the observed part of the distribution 
is therefore constant.

The exponential decay indicates that the decaying clusters are 
sampled from a strongly peaked part of the energy distribution,
rather than from a flat energy distribution.
To observe a single exponential decay for the total 
decay rate, the energy distribution should be more narrow than 
the function
\be
{\color{red}r(E) = k(E)\e^{-k(E)t},}
\ee
{\color{red}similar to the condition in Eq. \ref{gaussian0} to 
distinguish small and large photon energy scenarios.}
The expression for {\color{red} this} width is calculated in 
Eq.(\ref{width}).
With the values ${\rm EA} = 3.0$ eV \cite{Taylor1992},
$s = 3N-6$ valid for harmonic oscillators of a $N=5$ cluster,
and the frequency factor $\omega = 10^{14}$ s$^{-1}$ the value is
\be
\sigma_E = 0.045 ~{\rm eV}
\ee
The time scale, $t$, used in Eq.(\ref{width}) is the one for the 
measurement of the decay curves, which will summarily be set to 1 ms.

The energy emission rate reflected in the decrease of the emission
rate constant with time is calculated with the relation
\be
-4.96~ {\rm s}^{-1} = \frac{\d E}{\d t} \frac{\d \ln k}{\d E}.
\ee
A little algebra and setting the kinetic shift to zero, $E+E_0 
\approx {\rm EA}$, gives
\be
\label{emittedpower}
-\frac{\d E}{\d t} = 
\frac{4.96}{s-1}\frac{{\rm EA}}{\(\omega t\)^{1/(s-1)}} {\rm s}^{-1}
= 0.033 ~{\rm eV/s}.
\ee
This is an average value over the measured time, as it is 
based on the average logarithmic slope of the fitted rate 
constant (see Fig. \ref{Au5const}).

The small observed rate constant indicates that the energy
is emitted in vibrational transitions.
This is also consistent with the observation of purely exponential
decay rates fitted at all six laser firing times.
Larger photon energies would induce either loss of intensity,
which is not seen, as mentioned, or deviations from exponential 
decays, which is also not seen. 
In contrast to electronic transitions, vibrational transitions 
have a weak internal energy dependence.
As shown in ref. \cite{HansenJPCA2023}, this feature conserves 
the shape of energy distributions in an ensemble when it cools 
by emission of IR photons, apart from a scale factor, except at 
the lowest excitation energies.
Hence for such cases no width develops on the energy distribution 
with time, and the decay remains exponential, as observed.

\section{Summary and conclusions}

The spontaneous decays of anionic gold clusters were measured
in the DESIREE storage ring. 
The results showed strong similarities to the two other 
coinage metals, copper \cite{HansenPRA2017} and silver 
\cite{AndersonPRA2018}, in particular for the decay of the dimer, 
which was investigated with respect to the branching between electron 
and atom emission.

The radiative cooling times display a strong odd-even effect for
clusters $N=6-13$, similar to the behavior previously seen for the gold 
cationic clusters \cite{HansenPRA2017-2} but with the striking difference 
that anionic odd-$N$ clusters cool \textit{slower} than even-$N$, in a 
reversal of the oscillation pattern seen for the cations.
Furthermore, time constants are three to four orders longer for anions, 
but the relative
oscillation amplitude is quite similar for the two charge states. 
The fit of the decay curve of the heptamer indicates vibrational 
radiative cooling in this case. 
The precise reason for the very strong odd-even oscillations 
with size for gold clusters is still somewhat unclear.
Photo-excitation experiments on the pentamer showed that this cluster
size had cooled to a very low and well-defined internal energy already after 
0.12 s, making the decays exponential.
The decay constants were also found to vary exponentially with laser firing 
time, yielding an energy loss rate of 0.033 eV/s and a time constant of 0.20 s, 
indicating vibrational cooling also in this case.

In summary, the answers to the questions asked in the Introduction; 
'...whether the gold cluster anions will show RF and if they do, 
whether they will show the same odd-even
systematics as the cations' are probably not and definitely not, 
respectively.
The reason the first conclusion comes with a qualifier is that the 
strong indication of vibrational cooling is derived from the late 
photo-excitation experiments on cooled clusters, vs. the odd-even 
effect measured on the spontaneously decaying and therefore hotter 
clusters.

The other two questions both have clearer answers.
Yes, the behavior which we tentatively ascribe to
the presence of both high and low angular momenta is seen in the decay 
curves for the small clusters.
The gold dimer anion indeed shows a clear signature of breakdown of 
the Born-Oppenheimer separation, in close analogy to the silver and 
copper dimer anions. 

\section*{acknowledgement}
This work was performed at the Swedish National Research 
Infrastructure, DESIREE (Swedish Research Council Contract 
No. 2017-00621 and No. 2021-00155). 
It was supported with a beam time grant to KH.
HC, HZ, and HTS thank the Swedish Research Council 
for individual project grants (with Contracts No. 2023-03833, 
No. 2020-03437, and No. 2022-02822), and acknowledge the 
project grant "Probing charge- and mass- transfer reactions 
on the atomic level"  (2018.0028) from the Knut and Alice 
Wallenberg Foundation. 
This publication is based upon work from COST Action 
CA18212 - Molecular Dynamics in the GAS phase (MD-GAS), 
supported by COST (European Cooperation in Science and Technology).

\section{Appendix}

The fit of the spontaneous decay rates with power laws require that 
the time of creation of the clusters is know precisely. 
In this appendix we calculate the time offset and show that the 
value is close to zero with the preset electronic timing.
 
The counts in a single turn in the ring arise from ions 
that have been generated at different times in the source but 
stored equally long in the ring, modulo the spread in the decay 
position and the very small difference in the speed of the 
neutral detected.
The two spreads are so small that we can assign a single time to 
all decays that occur and are detected in any specific turn in the 
ring.
The binning of multiple turns improves statistics and has the 
added advantage that it tends to average out the already small but 
non-statistical betatron oscillations and related oscillations in the 
recorded neutralization yields that appear due to variations in 
detection efficencies \cite{MiyamotoPRST2008}.

The measured time of decay potentially differs from the 
time elapsed since creation of the clusters in the source,
as the acquisition timer is started during ion transit 
through the mass selection trajectory.
The time of detection refers to the neutral particle 
produced in a decay in the straight section of the ring 
in the detector side of the ring.  
Relative to the detection time, the decay time is earlier by
the amount of time it takes to move from the decay position 
to the detector, $t_{\rm dt}$.
It is
\be
t_{\rm dt} = 1.7 {\rm m}\sqrt{N} / v_1,
\ee
where the length is the flight distance between decay 
and detector and $v_1$ the speed of the monomer at the 
acceleration energy used for this mass.

We denote the number of turns in the ring before detection 
by $n$, starting the count with $n=1$ at the first half turn 
detection, corresponding to labeling the first peak seen 
in the spectrum as 1.
The time from production to detection of a decay during turn number 
$n$ is then given by $t_{\rm dt}$, by an electronic offset, 
$t_{\rm el}$ which includes both operation time of detector and a 
delay in acquisition electronics, the circulation time 
$t_{\rm ci}$, and the time from creation in the source to the 
arrival at the decay position in the first turn, $t_{\rm s}$.
For the detection time of peak $n$ the time is
\be
t_{N,n} =  t_{\rm el} + t_{\rm s} + (n-1)t_{\rm ci} + t_{\rm dt}.
\ee
The first term is cluster size independent. 
All other three terms are proportional to $\sqrt{N}$, up to
the two acceleration voltage used. 
In terms of the length of the ring, $\ell$, and the source-to-decay 
point length, $\ell_{\rm s}$,
the peak times are, with $v_N = v_1/\sqrt{N}$,
\be
t_{N,n} =  t_{\rm el} + \ell_{\rm s}/v_N + 
(n-1)\ell /v_N + 1.7 {\rm m}/ v_N .
\ee

The time resolved peaks for a given size allow us to plot $t_N$ vs. 
$n$ to find the parameter $\ell\sqrt{N}/v_1$ and the intercept 
given by $ t_{\rm el} + \ell_{\rm s} \sqrt{N}/v_1 + 
1.7 {\rm m}\sqrt{N}/ v_1$. 
Once these intercepts are determined, they are plotted 
vs. $\sqrt{N}$, from which we get the value for $\ell_{\rm s}/v_1 = 65.1$ 
$\mu$s from the slope.
The electronic offset $t_{\rm el}$ was found to be less 
than 1 $\mu$s, much smaller than the flight time of the
first detected peak, and can be ignored.
The value $\ell/v_1$ was found to 
be 86.9 $\mu$s, for the acceleration energy 10 keV.
The analysis then gives the true decay times in terms of 
the measured peak positions as
\be
t_{N,\rm true} =  t_N - 1.7 {\rm m}\sqrt{N}/ v_1.
\ee 


\begin{thebibliography}{34}
\expandafter\ifx\csname natexlab\endcsname\relax\def\natexlab#1{#1}\fi
\expandafter\ifx\csname bibnamefont\endcsname\relax
  \def\bibnamefont#1{#1}\fi
\expandafter\ifx\csname bibfnamefont\endcsname\relax
  \def\bibfnamefont#1{#1}\fi
\expandafter\ifx\csname citenamefont\endcsname\relax
  \def\citenamefont#1{#1}\fi
\expandafter\ifx\csname url\endcsname\relax
  \def\url#1{\texttt{#1}}\fi
\expandafter\ifx\csname urlprefix\endcsname\relax\def\urlprefix{URL }\fi
\providecommand{\bibinfo}[2]{#2}
\providecommand{\eprint}[2][]{\url{#2}}

\bibitem[{\citenamefont{Katakuse et~al.}(1985)\citenamefont{Katakuse, Ichihara,
  Fujita, Matsuo, Sakurai, and Matsuda}}]{Katakuse1985}
\bibinfo{author}{\bibfnamefont{I.}~\bibnamefont{Katakuse}},
  \bibinfo{author}{\bibfnamefont{T.}~\bibnamefont{Ichihara}},
  \bibinfo{author}{\bibfnamefont{Y.}~\bibnamefont{Fujita}},
  \bibinfo{author}{\bibfnamefont{T.}~\bibnamefont{Matsuo}},
  \bibinfo{author}{\bibfnamefont{T.}~\bibnamefont{Sakurai}}, \bibnamefont{and}
  \bibinfo{author}{\bibfnamefont{H.}~\bibnamefont{Matsuda}},
  \bibinfo{journal}{Int. J. Mass Spectrom. Ion Processes}
  \textbf{\bibinfo{volume}{67}}, \bibinfo{pages}{229–236}
  (\bibinfo{year}{1985}).

\bibitem[{\citenamefont{Katakuse et~al.}(1986)\citenamefont{Katakuse, Ichihara,
  Fujita, Matsuo, Sakurai, and Matsuda}}]{Katakuse1986}
\bibinfo{author}{\bibfnamefont{I.}~\bibnamefont{Katakuse}},
  \bibinfo{author}{\bibfnamefont{T.}~\bibnamefont{Ichihara}},
  \bibinfo{author}{\bibfnamefont{Y.}~\bibnamefont{Fujita}},
  \bibinfo{author}{\bibfnamefont{T.}~\bibnamefont{Matsuo}},
  \bibinfo{author}{\bibfnamefont{T.}~\bibnamefont{Sakurai}}, \bibnamefont{and}
  \bibinfo{author}{\bibfnamefont{H.}~\bibnamefont{Matsuda}},
  \bibinfo{journal}{Int. J. Mass Spectrom. Ion Proc.}
  \textbf{\bibinfo{volume}{74}}, \bibinfo{pages}{33} (\bibinfo{year}{1986}).

\bibitem[{\citenamefont{Hansen}(2020{\natexlab{a}})}]{HansenCP2020}
\bibinfo{author}{\bibfnamefont{K.}~\bibnamefont{Hansen}},
  \bibinfo{journal}{Chem. Phys.} \textbf{\bibinfo{volume}{530}},
  \bibinfo{pages}{110637(1} (\bibinfo{year}{2020}{\natexlab{a}}).

\bibitem[{\citenamefont{Nitzan and Jortner}(1979)}]{Nitzan1979}
\bibinfo{author}{\bibfnamefont{A.}~\bibnamefont{Nitzan}} \bibnamefont{and}
  \bibinfo{author}{\bibfnamefont{J.}~\bibnamefont{Jortner}},
  \bibinfo{journal}{J. Chem. Phys.} \textbf{\bibinfo{volume}{71}},
  \bibinfo{pages}{3524} (\bibinfo{year}{1979}).

\bibitem[{\citenamefont{Leach}(1987)}]{Leach1987}
\bibinfo{author}{\bibfnamefont{S.}~\bibnamefont{Leach}}, in
  \emph{\bibinfo{booktitle}{Polycyclic Aromatic Hydrocarbons and
  Astrophysics}}, edited by
  \bibinfo{editor}{\bibfnamefont{A.}~\bibnamefont{L{\'e}ger}},
  \bibinfo{editor}{\bibfnamefont{L.}~\bibnamefont{d'Hendecourt}},
  \bibnamefont{and} \bibinfo{editor}{\bibfnamefont{N.}~\bibnamefont{Boccara}}
  (\bibinfo{year}{1987}), vol. \bibinfo{volume}{191} of
  \emph{\bibinfo{series}{NATO ASI Series}}, pp. \bibinfo{pages}{99--127}, ISBN
  \bibinfo{isbn}{978-94-010-8619-6}.

\bibitem[{\citenamefont{L\'{e}ger et~al.}(1988)\citenamefont{L\'{e}ger,
  Boissel, and d'Hendecourt}}]{Leger1988}
\bibinfo{author}{\bibfnamefont{A.}~\bibnamefont{L\'{e}ger}},
  \bibinfo{author}{\bibfnamefont{P.}~\bibnamefont{Boissel}}, \bibnamefont{and}
  \bibinfo{author}{\bibfnamefont{L.}~\bibnamefont{d'Hendecourt}},
  \bibinfo{journal}{Phys. Rev. Lett.} \textbf{\bibinfo{volume}{60}},
  \bibinfo{pages}{921} (\bibinfo{year}{1988}).

\bibitem[{\citenamefont{Hansen et~al.}(2017{\natexlab{a}})\citenamefont{Hansen,
  Ferrari, Janssens, and Lievens}}]{HansenPRA2017-2}
\bibinfo{author}{\bibfnamefont{K.}~\bibnamefont{Hansen}},
  \bibinfo{author}{\bibfnamefont{P.}~\bibnamefont{Ferrari}},
  \bibinfo{author}{\bibfnamefont{E.}~\bibnamefont{Janssens}}, \bibnamefont{and}
  \bibinfo{author}{\bibfnamefont{P.}~\bibnamefont{Lievens}},
  \bibinfo{journal}{Phys. Rev. A} \textbf{\bibinfo{volume}{96}},
  \bibinfo{pages}{022511} (\bibinfo{year}{2017}{\natexlab{a}}).

\bibitem[{\citenamefont{Green et~al.}(2021)\citenamefont{Green, Gentleman,
  Sch\"{o}llkopf, Fielicke, and Mackenzie}}]{GreenPRL2021}
\bibinfo{author}{\bibfnamefont{A.~E.} \bibnamefont{Green}},
  \bibinfo{author}{\bibfnamefont{A.~S.} \bibnamefont{Gentleman}},
  \bibinfo{author}{\bibfnamefont{W.}~\bibnamefont{Sch\"{o}llkopf}},
  \bibinfo{author}{\bibfnamefont{A.}~\bibnamefont{Fielicke}}, \bibnamefont{and}
  \bibinfo{author}{\bibfnamefont{S.~R.} \bibnamefont{Mackenzie}},
  \bibinfo{journal}{Phys. Rev. Lett.} \textbf{\bibinfo{volume}{127}},
  \bibinfo{pages}{033002} (\bibinfo{year}{2021}).

\bibitem[{\citenamefont{Hansen et~al.}(2017{\natexlab{b}})\citenamefont{Hansen,
  Ferrari~Ramirez, Janssens, and Lievens}}]{HansenJPCC2017}
\bibinfo{author}{\bibfnamefont{K.}~\bibnamefont{Hansen}},
  \bibinfo{author}{\bibfnamefont{P.}~\bibnamefont{Ferrari~Ramirez}},
  \bibinfo{author}{\bibfnamefont{E.}~\bibnamefont{Janssens}}, \bibnamefont{and}
  \bibinfo{author}{\bibfnamefont{P.}~\bibnamefont{Lievens}},
  \bibinfo{journal}{J. Phys. Chem. C} \textbf{\bibinfo{volume}{121}},
  \bibinfo{pages}{10663} (\bibinfo{year}{2017}{\natexlab{b}}).

\bibitem[{\citenamefont{Jackschath et~al.}(1992)\citenamefont{Jackschath,
  Rabin, , and Schulze}}]{JackschathBBPC1992}
\bibinfo{author}{\bibfnamefont{C.}~\bibnamefont{Jackschath}},
  \bibinfo{author}{\bibfnamefont{I.}~\bibnamefont{Rabin}}, , \bibnamefont{and}
  \bibinfo{author}{\bibfnamefont{W.}~\bibnamefont{Schulze}},
  \bibinfo{journal}{Ber. Bunsenges. Phys. Chem} \textbf{\bibinfo{volume}{96}},
  \bibinfo{pages}{1200} (\bibinfo{year}{1992}).

\bibitem[{\citenamefont{Ho et~al.}(1990)\citenamefont{Ho, Ervin, and
  Lineberger}}]{Ho90}
\bibinfo{author}{\bibfnamefont{J.}~\bibnamefont{Ho}},
  \bibinfo{author}{\bibfnamefont{K.~M.} \bibnamefont{Ervin}}, \bibnamefont{and}
  \bibinfo{author}{\bibfnamefont{W.~C.} \bibnamefont{Lineberger}},
  \bibinfo{journal}{J. Chem. Phys.} \textbf{\bibinfo{volume}{93}},
  \bibinfo{pages}{6987} (\bibinfo{year}{1990}).

\bibitem[{\citenamefont{Taylor et~al.}(1992)\citenamefont{Taylor,
  Pettiette-Hall, Cheshnovsky, and Smalley}}]{Taylor1992}
\bibinfo{author}{\bibfnamefont{K.~J.} \bibnamefont{Taylor}},
  \bibinfo{author}{\bibfnamefont{C.~L.} \bibnamefont{Pettiette-Hall}},
  \bibinfo{author}{\bibfnamefont{O.}~\bibnamefont{Cheshnovsky}},
  \bibnamefont{and} \bibinfo{author}{\bibfnamefont{R.~E.}
  \bibnamefont{Smalley}}, \bibinfo{journal}{J. Chem. Phys.}
  \textbf{\bibinfo{volume}{96}}, \bibinfo{pages}{3319} (\bibinfo{year}{1992}),
  \urlprefix\url{http://scitation.aip.org/content/aip/journal/jcp/96/4/10.1063/1.461927}.

\bibitem[{\citenamefont{H{\"a}kkinen et~al.}(2003)\citenamefont{H{\"a}kkinen,
  Yoon, Landman, Li, Zhai, and Wang}}]{HakkinenJPCA2003}
\bibinfo{author}{\bibfnamefont{H.}~\bibnamefont{H{\"a}kkinen}},
  \bibinfo{author}{\bibfnamefont{B.}~\bibnamefont{Yoon}},
  \bibinfo{author}{\bibfnamefont{U.}~\bibnamefont{Landman}},
  \bibinfo{author}{\bibfnamefont{X.}~\bibnamefont{Li}},
  \bibinfo{author}{\bibfnamefont{H.-J.} \bibnamefont{Zhai}}, \bibnamefont{and}
  \bibinfo{author}{\bibfnamefont{L.-S.} \bibnamefont{Wang}},
  \bibinfo{journal}{J. Phys. Chem. A} \textbf{\bibinfo{volume}{107}},
  \bibinfo{pages}{6168} (\bibinfo{year}{2003}).

\bibitem[{\citenamefont{Vogel et~al.}(2001)\citenamefont{Vogel, Hansen,
  Herlert, and Schweikhard}}]{VogelPRL2001}
\bibinfo{author}{\bibfnamefont{M.}~\bibnamefont{Vogel}},
  \bibinfo{author}{\bibfnamefont{K.}~\bibnamefont{Hansen}},
  \bibinfo{author}{\bibfnamefont{A.}~\bibnamefont{Herlert}}, \bibnamefont{and}
  \bibinfo{author}{\bibfnamefont{L.}~\bibnamefont{Schweikhard}},
  \bibinfo{journal}{Phys. Rev. Lett.} \textbf{\bibinfo{volume}{87}},
  \bibinfo{pages}{013401} (\bibinfo{year}{2001}).

\bibitem[{\citenamefont{Hansen et~al.}(2006)\citenamefont{Hansen, Herlert,
  Schweikhard, and Vogel}}]{HansenPRA2006}
\bibinfo{author}{\bibfnamefont{K.}~\bibnamefont{Hansen}},
  \bibinfo{author}{\bibfnamefont{A.}~\bibnamefont{Herlert}},
  \bibinfo{author}{\bibfnamefont{L.}~\bibnamefont{Schweikhard}},
  \bibnamefont{and} \bibinfo{author}{\bibfnamefont{M.}~\bibnamefont{Vogel}},
  \bibinfo{journal}{Phys. Rev. A} \textbf{\bibinfo{volume}{73}},
  \bibinfo{pages}{063202} (\bibinfo{year}{2006}).

\bibitem[{\citenamefont{Hansen et~al.}(2017{\natexlab{c}})\citenamefont{Hansen,
  Stockett, Kaminska, Nascimento, Anderson, Gatchell, Chartkunchand, Eklund,
  Zettergren, Schmidt et~al.}}]{HansenPRA2017}
\bibinfo{author}{\bibfnamefont{K.}~\bibnamefont{Hansen}},
  \bibinfo{author}{\bibfnamefont{M.~H.} \bibnamefont{Stockett}},
  \bibinfo{author}{\bibfnamefont{M.}~\bibnamefont{Kaminska}},
  \bibinfo{author}{\bibfnamefont{R.~F.} \bibnamefont{Nascimento}},
  \bibinfo{author}{\bibfnamefont{E.~K.} \bibnamefont{Anderson}},
  \bibinfo{author}{\bibfnamefont{M.}~\bibnamefont{Gatchell}},
  \bibinfo{author}{\bibfnamefont{K.~C.} \bibnamefont{Chartkunchand}},
  \bibinfo{author}{\bibfnamefont{G.}~\bibnamefont{Eklund}},
  \bibinfo{author}{\bibfnamefont{H.}~\bibnamefont{Zettergren}},
  \bibinfo{author}{\bibfnamefont{H.~T.} \bibnamefont{Schmidt}},
  \bibnamefont{et~al.}, \bibinfo{journal}{Phys. Rev. A}
  \textbf{\bibinfo{volume}{95}}, \bibinfo{pages}{022511}
  (\bibinfo{year}{2017}{\natexlab{c}}).

\bibitem[{\citenamefont{Anderson et~al.}(2018)\citenamefont{Anderson,
  Kami{\'n}ska, Chartkunchand, Eklund, Gatchell, Hansen, Zettergren,
  Cederquist, and Schmidt}}]{AndersonPRA2018}
\bibinfo{author}{\bibfnamefont{E.~K.} \bibnamefont{Anderson}},
  \bibinfo{author}{\bibfnamefont{M.}~\bibnamefont{Kami{\'n}ska}},
  \bibinfo{author}{\bibfnamefont{K.~C.} \bibnamefont{Chartkunchand}},
  \bibinfo{author}{\bibfnamefont{G.}~\bibnamefont{Eklund}},
  \bibinfo{author}{\bibfnamefont{M.}~\bibnamefont{Gatchell}},
  \bibinfo{author}{\bibfnamefont{K.}~\bibnamefont{Hansen}},
  \bibinfo{author}{\bibfnamefont{H.}~\bibnamefont{Zettergren}},
  \bibinfo{author}{\bibfnamefont{H.}~\bibnamefont{Cederquist}},
  \bibnamefont{and} \bibinfo{author}{\bibfnamefont{H.~T.}
  \bibnamefont{Schmidt}}, \bibinfo{journal}{Phys. Rev. A}
  \textbf{\bibinfo{volume}{98}}, \bibinfo{pages}{022705}
  (\bibinfo{year}{2018}).

\bibitem[{\citenamefont{Anderson et~al.}(2020)\citenamefont{Anderson,
  Schmidt-May, Najeeb, Eklund, Chartkunchand, Ros{\'e}n, Larson, Hansen,
  Cederquist, Zettergren et~al.}}]{AndersonPRL2020}
\bibinfo{author}{\bibfnamefont{E.~K.} \bibnamefont{Anderson}},
  \bibinfo{author}{\bibfnamefont{A.~F.} \bibnamefont{Schmidt-May}},
  \bibinfo{author}{\bibfnamefont{P.~K.} \bibnamefont{Najeeb}},
  \bibinfo{author}{\bibfnamefont{G.}~\bibnamefont{Eklund}},
  \bibinfo{author}{\bibfnamefont{K.~C.} \bibnamefont{Chartkunchand}},
  \bibinfo{author}{\bibfnamefont{S.}~\bibnamefont{Ros{\'e}n}},
  \bibinfo{author}{\bibfnamefont{{\AA}.}~\bibnamefont{Larson}},
  \bibinfo{author}{\bibfnamefont{K.}~\bibnamefont{Hansen}},
  \bibinfo{author}{\bibfnamefont{H.}~\bibnamefont{Cederquist}},
  \bibinfo{author}{\bibfnamefont{H.}~\bibnamefont{Zettergren}},
  \bibnamefont{et~al.}, \bibinfo{journal}{Phys. Rev. Lett.}
  \textbf{\bibinfo{volume}{124}}, \bibinfo{pages}{173001}
  (\bibinfo{year}{2020}).

\bibitem[{\citenamefont{Anderson et~al.}(2023)\citenamefont{Anderson,
  Schmidt-May, Najeeb, Eklund, Chartkunchand, Ros{\'e}n, Kami{\'n}ska,
  Stockett, Nascimento, Hansen et~al.}}]{AndersonPRA2023}
\bibinfo{author}{\bibfnamefont{E.~K.} \bibnamefont{Anderson}},
  \bibinfo{author}{\bibfnamefont{A.~F.} \bibnamefont{Schmidt-May}},
  \bibinfo{author}{\bibfnamefont{P.~K.} \bibnamefont{Najeeb}},
  \bibinfo{author}{\bibfnamefont{G.}~\bibnamefont{Eklund}},
  \bibinfo{author}{\bibfnamefont{K.~C.} \bibnamefont{Chartkunchand}},
  \bibinfo{author}{\bibfnamefont{S.}~\bibnamefont{Ros{\'e}n}},
  \bibinfo{author}{\bibfnamefont{M.}~\bibnamefont{Kami{\'n}ska}},
  \bibinfo{author}{\bibfnamefont{M.~H.} \bibnamefont{Stockett}},
  \bibinfo{author}{\bibfnamefont{R.}~\bibnamefont{Nascimento}},
  \bibinfo{author}{\bibfnamefont{K.}~\bibnamefont{Hansen}},
  \bibnamefont{et~al.}, \bibinfo{journal}{Phys. Rev. A}
  \textbf{\bibinfo{volume}{107}}, \bibinfo{pages}{062824}
  (\bibinfo{year}{2023}).

\bibitem[{\citenamefont{Thomas et~al.}(2011)\citenamefont{Thomas, Schmidt,
  Andler, Bj\"{o}rkhage, Blom, Br\"{a}nnholm, B\"{a}ckstr\"{o}m, Danared, Das,
  Haag et~al.}}]{ThomasRSI2011}
\bibinfo{author}{\bibfnamefont{R.~D.} \bibnamefont{Thomas}},
  \bibinfo{author}{\bibfnamefont{H.~T.} \bibnamefont{Schmidt}},
  \bibinfo{author}{\bibfnamefont{G.}~\bibnamefont{Andler}},
  \bibinfo{author}{\bibfnamefont{M.}~\bibnamefont{Bj\"{o}rkhage}},
  \bibinfo{author}{\bibfnamefont{M.}~\bibnamefont{Blom}},
  \bibinfo{author}{\bibfnamefont{L.}~\bibnamefont{Br\"{a}nnholm}},
  \bibinfo{author}{\bibfnamefont{E.}~\bibnamefont{B\"{a}ckstr\"{o}m}},
  \bibinfo{author}{\bibfnamefont{H.}~\bibnamefont{Danared}},
  \bibinfo{author}{\bibfnamefont{S.}~\bibnamefont{Das}},
  \bibinfo{author}{\bibfnamefont{N.}~\bibnamefont{Haag}}, \bibnamefont{et~al.},
  \bibinfo{journal}{Review of Scientific Instruments}
  \textbf{\bibinfo{volume}{82}}, \bibinfo{pages}{065112}
  (\bibinfo{year}{2011}).

\bibitem[{\citenamefont{Schmidt et~al.}(2013)\citenamefont{Schmidt, Thomas,
  Gatchell, Ros{\'e}n, Reinhed, L{\"o}fgren, Br{\"a}nnholm, Blom,
  Bj{\"o}rkhage, B{\"a}ckstr{\"o}m et~al.}}]{SchmidtRSI2013}
\bibinfo{author}{\bibfnamefont{H.~T.} \bibnamefont{Schmidt}},
  \bibinfo{author}{\bibfnamefont{R.~D.} \bibnamefont{Thomas}},
  \bibinfo{author}{\bibfnamefont{M.}~\bibnamefont{Gatchell}},
  \bibinfo{author}{\bibfnamefont{S.}~\bibnamefont{Ros{\'e}n}},
  \bibinfo{author}{\bibfnamefont{P.}~\bibnamefont{Reinhed}},
  \bibinfo{author}{\bibfnamefont{P.}~\bibnamefont{L{\"o}fgren}},
  \bibinfo{author}{\bibfnamefont{L.}~\bibnamefont{Br{\"a}nnholm}},
  \bibinfo{author}{\bibfnamefont{M.}~\bibnamefont{Blom}},
  \bibinfo{author}{\bibfnamefont{M.}~\bibnamefont{Bj{\"o}rkhage}},
  \bibinfo{author}{\bibfnamefont{E.}~\bibnamefont{B{\"a}ckstr{\"o}m}},
  \bibnamefont{et~al.}, \bibinfo{journal}{Rev. Sci. Instrum.}
  \textbf{\bibinfo{volume}{84}}, \bibinfo{pages}{055115}
  (\bibinfo{year}{2013}).

\bibitem[{\citenamefont{Taylor et~al.}(1990)\citenamefont{Taylor, Jin,
  Conceicao, Wang, Cheshnovsky, Johnson, Nordlander, and
  Smalley}}]{TaylorJCP1990}
\bibinfo{author}{\bibfnamefont{K.~J.} \bibnamefont{Taylor}},
  \bibinfo{author}{\bibfnamefont{C.}~\bibnamefont{Jin}},
  \bibinfo{author}{\bibfnamefont{J.}~\bibnamefont{Conceicao}},
  \bibinfo{author}{\bibfnamefont{L.}~\bibnamefont{Wang}},
  \bibinfo{author}{\bibfnamefont{O.}~\bibnamefont{Cheshnovsky}},
  \bibinfo{author}{\bibfnamefont{B.~R.} \bibnamefont{Johnson}},
  \bibinfo{author}{\bibfnamefont{P.~J.} \bibnamefont{Nordlander}},
  \bibnamefont{and} \bibinfo{author}{\bibfnamefont{R.~E.}
  \bibnamefont{Smalley}}, \bibinfo{journal}{J. Chem. Phys.}
  \textbf{\bibinfo{volume}{93}}, \bibinfo{pages}{7515} (\bibinfo{year}{1990}).

\bibitem[{\citenamefont{B\"ackstr\"om et~al.}(2015)\citenamefont{B\"ackstr\"om,
  Hanstorp, Hole, Kaminska, Nascimento, Blom, Bj\"orkhage, K\"allberg,
  L\"ofgren, Reinhed et~al.}}]{BackstromPRL2015}
\bibinfo{author}{\bibfnamefont{E.}~\bibnamefont{B\"ackstr\"om}},
  \bibinfo{author}{\bibfnamefont{D.}~\bibnamefont{Hanstorp}},
  \bibinfo{author}{\bibfnamefont{O.~M.} \bibnamefont{Hole}},
  \bibinfo{author}{\bibfnamefont{M.}~\bibnamefont{Kaminska}},
  \bibinfo{author}{\bibfnamefont{R.~F.} \bibnamefont{Nascimento}},
  \bibinfo{author}{\bibfnamefont{M.}~\bibnamefont{Blom}},
  \bibinfo{author}{\bibfnamefont{M.}~\bibnamefont{Bj\"orkhage}},
  \bibinfo{author}{\bibfnamefont{A.}~\bibnamefont{K\"allberg}},
  \bibinfo{author}{\bibfnamefont{P.}~\bibnamefont{L\"ofgren}},
  \bibinfo{author}{\bibfnamefont{P.}~\bibnamefont{Reinhed}},
  \bibnamefont{et~al.}, \bibinfo{journal}{Phys. Rev. Lett.}
  \textbf{\bibinfo{volume}{114}}, \bibinfo{pages}{143003}
  (\bibinfo{year}{2015}),
  \urlprefix\url{https://link.aps.org/doi/10.1103/PhysRevLett.114.143003}.

\bibitem[{\citenamefont{Hansen et~al.}(2001)\citenamefont{Hansen, Andersen,
  Hvelplund, M{\o}ller, Pedersen, and Petrunin}}]{HansenPRL2001}
\bibinfo{author}{\bibfnamefont{K.}~\bibnamefont{Hansen}},
  \bibinfo{author}{\bibfnamefont{J.~U.} \bibnamefont{Andersen}},
  \bibinfo{author}{\bibfnamefont{P.}~\bibnamefont{Hvelplund}},
  \bibinfo{author}{\bibfnamefont{S.~P.} \bibnamefont{M{\o}ller}},
  \bibinfo{author}{\bibfnamefont{U.~V.} \bibnamefont{Pedersen}},
  \bibnamefont{and} \bibinfo{author}{\bibfnamefont{V.~V.}
  \bibnamefont{Petrunin}}, \bibinfo{journal}{Phys. Rev. Lett.}
  \textbf{\bibinfo{volume}{87}}, \bibinfo{pages}{123401}
  (\bibinfo{year}{2001}).

\bibitem[{\citenamefont{Fedor et~al.}(2005)\citenamefont{Fedor, Hansen,
  Andersen, and Hvelplund}}]{fedor05}
\bibinfo{author}{\bibfnamefont{J.}~\bibnamefont{Fedor}},
  \bibinfo{author}{\bibfnamefont{K.}~\bibnamefont{Hansen}},
  \bibinfo{author}{\bibfnamefont{J.~U.} \bibnamefont{Andersen}},
  \bibnamefont{and}
  \bibinfo{author}{\bibfnamefont{P.}~\bibnamefont{Hvelplund}},
  \bibinfo{journal}{Phys. Rev. Lett.} \textbf{\bibinfo{volume}{94}},
  \bibinfo{pages}{113201} (\bibinfo{year}{2005}).

\bibitem[{\citenamefont{Menk et~al.}(2014)\citenamefont{Menk, Das, Blaum,
  Froese, Lange, Mukherjee, Repnow, Schwalm, von Hahn, and Wolf}}]{MenkPRA2014}
\bibinfo{author}{\bibfnamefont{S.}~\bibnamefont{Menk}},
  \bibinfo{author}{\bibfnamefont{S.}~\bibnamefont{Das}},
  \bibinfo{author}{\bibfnamefont{K.}~\bibnamefont{Blaum}},
  \bibinfo{author}{\bibfnamefont{M.~W.} \bibnamefont{Froese}},
  \bibinfo{author}{\bibfnamefont{M.}~\bibnamefont{Lange}},
  \bibinfo{author}{\bibfnamefont{M.}~\bibnamefont{Mukherjee}},
  \bibinfo{author}{\bibfnamefont{R.}~\bibnamefont{Repnow}},
  \bibinfo{author}{\bibfnamefont{D.}~\bibnamefont{Schwalm}},
  \bibinfo{author}{\bibfnamefont{R.}~\bibnamefont{von Hahn}}, \bibnamefont{and}
  \bibinfo{author}{\bibfnamefont{A.}~\bibnamefont{Wolf}},
  \bibinfo{journal}{Phys. Rev. A} \textbf{\bibinfo{volume}{89}},
  \bibinfo{pages}{022502} (\bibinfo{year}{2014}).

\bibitem[{\citenamefont{Ferrari et~al.}(2019)\citenamefont{Ferrari, Janssens,
  Lievens, and Hansen}}]{FerrariIRPC2019}
\bibinfo{author}{\bibfnamefont{P.}~\bibnamefont{Ferrari}},
  \bibinfo{author}{\bibfnamefont{E.}~\bibnamefont{Janssens}},
  \bibinfo{author}{\bibfnamefont{P.}~\bibnamefont{Lievens}}, \bibnamefont{and}
  \bibinfo{author}{\bibfnamefont{K.}~\bibnamefont{Hansen}},
  \bibinfo{journal}{Int. Rev. Phys. Chem.} \textbf{\bibinfo{volume}{38}},
  \bibinfo{pages}{405} (\bibinfo{year}{2019}).

\bibitem[{\citenamefont{Hansen}(2018)}]{book}
\bibinfo{author}{\bibfnamefont{K.}~\bibnamefont{Hansen}},
  \emph{\bibinfo{title}{Statistical Physics of Nanoparticles in the Gas
  Phase}}, vol.~\bibinfo{volume}{73} of \emph{\bibinfo{series}{Springer Series
  on Atomic, Optical, and Plasma Physics}} (\bibinfo{publisher}{Springer},
  \bibinfo{address}{Dordrecht}, \bibinfo{year}{2018}), ISBN
  \bibinfo{isbn}{978-3-319-90061-2}.

\bibitem[{\citenamefont{Andersen et~al.}(2001)\citenamefont{Andersen, Bonderup,
  and Hansen}}]{AndersenJCP2001}
\bibinfo{author}{\bibfnamefont{J.~U.} \bibnamefont{Andersen}},
  \bibinfo{author}{\bibfnamefont{E.}~\bibnamefont{Bonderup}}, \bibnamefont{and}
  \bibinfo{author}{\bibfnamefont{K.}~\bibnamefont{Hansen}},
  \bibinfo{journal}{J. Chem. Phys.} \textbf{\bibinfo{volume}{114}},
  \bibinfo{pages}{6518} (\bibinfo{year}{2001}).

\bibitem[{\citenamefont{Hansen and Campbell}(1996)}]{HansenJCP1996}
\bibinfo{author}{\bibfnamefont{K.}~\bibnamefont{Hansen}} \bibnamefont{and}
  \bibinfo{author}{\bibfnamefont{E.~E.~B.} \bibnamefont{Campbell}},
  \bibinfo{journal}{J. Chem. Phys.} \textbf{\bibinfo{volume}{104}},
  \bibinfo{pages}{5012} (\bibinfo{year}{1996}),
  \urlprefix\url{http://scitation.aip.org/content/aip/journal/jcp/104/13/10.1063/1.471130}.

\bibitem[{\citenamefont{Andersen et~al.}(1996)\citenamefont{Andersen, Brink,
  Hvelplund, Larsson, Nielsen, and Shen}}]{andersen1996}
\bibinfo{author}{\bibfnamefont{J.~U.} \bibnamefont{Andersen}},
  \bibinfo{author}{\bibfnamefont{C.}~\bibnamefont{Brink}},
  \bibinfo{author}{\bibfnamefont{P.}~\bibnamefont{Hvelplund}},
  \bibinfo{author}{\bibfnamefont{M.~O.} \bibnamefont{Larsson}},
  \bibinfo{author}{\bibfnamefont{B.~B.} \bibnamefont{Nielsen}},
  \bibnamefont{and} \bibinfo{author}{\bibfnamefont{H.}~\bibnamefont{Shen}},
  \bibinfo{journal}{Phys. Rev. Lett.} \textbf{\bibinfo{volume}{77}},
  \bibinfo{pages}{3991} (\bibinfo{year}{1996}).

\bibitem[{\citenamefont{Hansen}(2020{\natexlab{b}})}]{HansenPRA2020}
\bibinfo{author}{\bibfnamefont{K.}~\bibnamefont{Hansen}},
  \bibinfo{journal}{Phys. Rev. A} \textbf{\bibinfo{volume}{102}},
  \bibinfo{pages}{052823} (\bibinfo{year}{2020}{\natexlab{b}}).

\bibitem[{\citenamefont{Hansen et~al.}(2023)\citenamefont{Hansen, Licht,
  Kurbanov, and Toker}}]{HansenJPCA2023}
\bibinfo{author}{\bibfnamefont{K.}~\bibnamefont{Hansen}},
  \bibinfo{author}{\bibfnamefont{O.}~\bibnamefont{Licht}},
  \bibinfo{author}{\bibfnamefont{A.}~\bibnamefont{Kurbanov}}, \bibnamefont{and}
  \bibinfo{author}{\bibfnamefont{Y.}~\bibnamefont{Toker}}, \bibinfo{journal}{J.
  Phys. Chem. A} \textbf{\bibinfo{volume}{127}}, \bibinfo{pages}{2889}
  (\bibinfo{year}{2023}).

\bibitem[{\citenamefont{Miyamoto et~al.}(2008)\citenamefont{Miyamoto, Kopp,
  Jansson, and Syphers}}]{MiyamotoPRST2008}
\bibinfo{author}{\bibfnamefont{R.}~\bibnamefont{Miyamoto}},
  \bibinfo{author}{\bibfnamefont{S.~E.} \bibnamefont{Kopp}},
  \bibinfo{author}{\bibfnamefont{A.}~\bibnamefont{Jansson}}, \bibnamefont{and}
  \bibinfo{author}{\bibfnamefont{M.~J.} \bibnamefont{Syphers}},
  \bibinfo{journal}{Phys. Rev. ST Accel. Beams} \textbf{\bibinfo{volume}{11}},
  \bibinfo{pages}{084002} (\bibinfo{year}{2008}).

\end{thebibliography}

\end{document}